\documentclass[conference]{IEEEtran}
\IEEEoverridecommandlockouts

\usepackage{amsmath,amssymb,amsfonts,mathtools,mathrsfs,cite,graphicx,color}
\usepackage{fullpage}
\usepackage[colorlinks=true,urlcolor=blue]{hyperref}
\usepackage{algorithmic}
\usepackage{textcomp}
\usepackage{xcolor}
\def\BibTeX{{\rm B\kern-.05em{\sc i\kern-.025em b}\kern-.08em
    T\kern-.1667em\lower.7ex\hbox{E}\kern-.125emX}}

\def \endprf{\hfill {\vrule height6pt width6pt depth0pt}\medskip}

\newcommand{\R}{\mathbb{R}}

\definecolor{chartreuse}{RGB}{110,220,0}

\renewcommand{\Re}[1]{\operatorname{Re}\left\{#1\right\}}
\renewcommand{\Im}[1]{\operatorname{Im}\left\{#1\right\}}

\newcommand{\vct}[1]{\boldsymbol{#1}}
\newcommand{\mtx}[1]{\boldsymbol{#1}}

\newcommand{\T}{\mathrm{T}}

\DeclareMathOperator*{\argmin}{\text{arg~min}}

\newcommand{\vk}{\vct{k}}

\newcommand{\vr}{\vct{r}}

\newcommand{\vx}{\vct{x}}

\newcommand{\valpha}{\vct{\alpha}}

\newcommand{\veta}{\vct{\eta}}
\newcommand{\vtheta}{\vct{\theta}}

\newcommand{\mH}{\mtx{H}}

\newcommand{\mU}{\mtx{U}}

\newcommand{\mX}{\mtx{X}}
\newcommand{\mY}{\mtx{Y}}
\newcommand{\mZ}{\mtx{Z}}
\newcommand{\mId}{{\bf I}}

\newcommand{\mzero}{{\bf 0}}

\begin{document}


\title{Subspace Tracking with Dynamical Models on the Grassmannian
\thanks{This work was supported in part by COGNISENSE, one of seven centers in JUMP 2.0, a Semiconductor Research Corporation (SRC) program sponsored by DARPA.}
}

\author{\IEEEauthorblockN{Alex Saad-Falcon, Brighton Ancelin, and Justin Romberg}
\IEEEauthorblockA{\textit{Department of Electrical and Computer Engineering} \\
\textit{Georgia Institute of Technology}\\
Atlanta, GA, USA \\
alexsaadfalcon@gatech.edu}
}

\maketitle

\begin{abstract}
Tracking signals in dynamic environments presents difficulties in both analysis and implementation. In this work, we expand on a class of subspace tracking algorithms which utilize the Grassmann manifold -- the set of linear subspaces of a high-dimensional vector space. We design regularized least squares algorithms based on common manifold operations and intuitive dynamical models. We demonstrate the efficacy of the approach for a narrowband beamforming scenario, where the dynamics of multiple signals of interest are captured by motion on the Grassmannian.
\end{abstract}


\vspace*{-5pt}
\section{Introduction}
\vspace*{-5pt}

Representing latent structure in high dimensional data is often accomplished through low-dimensional linear subspaces, e.g., principal component analysis (PCA). For high-dimensional data or large datasets, algorithms have been designed for ``streaming" or ``online" PCA, where data are observed sample by sample, and the subspace estimate is progressively updated \cite{bin_yang_projection_1995,jain_streaming_2016,henriksen_adaoja_2019,balzano_equivalence_2022}. Numerous algorithms and variations have been developed which address different settings, e.g., for partially or compressively observed data \cite{balzano_online_2011,chi_petrels_2013,balzano_streaming_2018}, suddenly changing subspaces \cite{narayanamurthy_provable_subspace_2019}, or data subject to outliers \cite{vaswani_robust_2018}. A recent class of algorithms incorporates the topology of the Grassmann manifold, also called the Grassmannian \cite{balzano_online_2011,zhang_global_2016,zhang_convergence_2022_2,blocker_dynamic_2023}.

In this work, we consider a setting similar to the work in \cite{blocker_dynamic_2023} in which we have access to batches of full-dimensional data which we aim to fit with a time-evolving subspace on the Grassmannian. In contrast to previous work, we do not fit a single geodesic or even piecewise geodesics to the data. We estimate a subspace for each batch while incentivizing the subspaces to trace a smooth trajectory on the Grassmannian through regularization.
The additional structure provided by the smooth trajectory -- for which ``smoothness" is defined explicitly -- enables accurate subspace tracking from noisy measurements.
We approximate a few Grassmannian operations to improve computational efficiency with minimal performance penalty. 


Our motivation for developing this algorithmic framework stems from antenna array processing. Bandlimited signals impinging on an array are well-represented by low-dimensional subspaces \cite{delude_broadband_2022}.
However, multiple signals with varying angles of arrival (AoA) cause changing subspaces.
The emergent dynamics of subspace motion can be very complicated; small changes in the AoA of various signals can result in significant subspace motion and even changes in rank. We aim to encode complex signal dynamics with smooth motion on the Grassmannian. The manifold structure enables us to relate subspaces and express subspace motion using concepts analogous to particle position and velocity in Euclidean space.


\vspace*{-5pt}
\section{Formulation}\label{sec:formulation}
\vspace*{-5pt}

The Grassmannian, denoted by the set $G(n, d)$, is the set of all $d$-dimensional linear subspaces of an $n$-dimensional vector space. Subspace estimation and tracking can be formulated as algorithms where iterates are constrained to be elements of $G(n, d)$. As a quotient manifold, the Grassmannian allows us to express the natural invariance to choice of basis of a subspace while estimating/tracking. Refer to \cite{boumal_introduction_2023} for a cohesive guide to differential geometry and optimization on manifolds.


Points on the Grassmannian require analogous operations to those for points in Euclidean space. A point on the Grassmannian $\mathcal{Y}\in G(n,d)$ can by modeled by a representative basis for the subspace $\mY\in\R^{n\times d},\mY^T\mY=\mId$ (while keeping in mind that this choice is not unique). Comparisons between points $\mY, \mZ \in G(n,d)$ are often described in terms of the \textit{principal angles} $\left\{\theta_i\right\}_{i=1}^d \in \left[0, \pi/2\right]^d$ between them (by convention $\theta_i \leq \theta_{i+1}$). The principal angles can be computed as
$\theta_i (\mY, \mZ) = \arccos(\sigma_i(\mY^T\mZ))$ for singular values $\sigma_i$.

Two notions of distance on the Grassmannian are the \textit{geodesic} (or \textit{arc length}) and \textit{chordal} (or \textit{projection F-norm}) distances, defined as $||\vtheta||_2$ and $||\sin(\vtheta)||_2 = 2^{-1/2} ||\mZ \mZ^\T - \mY \mY^T||_F$ respectively \cite{edelman_geometry_1998}. The definitions are equivalent up to first order. Geodesic distance is in some sense the ``natural" metric (akin to $\ell^2$ norm from Euclidean geometry). However, using it in objective functions can lead to less efficient optimization algorithms. Computing the geodesic distance requires a singular value decomposition (SVD), whereas the chordal distance only requires matrix multiplies and norms.

In addition to a distance operation, we also require operations for motion on the manifold. Each point on the manifold $\mY\in G(n,d)$ is equipped with a \textit{tangent space} $T_{\mY} G(n,d)$ -- a set of travel directions from $\mY$ which locally stay on the manifold. The tangent space of a point on the Grassmannian $\mY$ is the set $\{\mH\in\R^{n\times d} : \mY^T\mH=\mzero\}$.
Moving from $\mY$ in any tangent direction will eventually leave the manifold, so the \textit{exponential map} is used to travel along the manifold. $\exp_{\mY}(\mH)=\mZ$ takes a point and tangent vector and produces a new point $\mZ\in G(n,d)$ which is the result of following a geodesic in the tangent direction.
The exponential map on the Grassmannian is computed in closed form using the SVD of $\mH$ \cite{Bendokat_2024} and represents a mixing between the original basis $\mY$ and the new basis $\rm{orth}(\mH)$.

The \textit{logarithmic map} $\log_{\mY}(\mZ)=\mH$ is the inverse of the exponential map if $\mY$ and $\mZ$ are close on the Grassmannian ($\theta_d<\pi/2$, within the injectivity radius \cite{Bendokat_2024}) and relates to geodesic distance ($\|\log_{\mY}(\mZ)\|_F=\|\vtheta\|_2$). The logarithmic map gives us the direction of travel $\mH$ which takes us on the shortest path from $\mY$ to $\mZ$,
which intuitively represents the change needed to convert basis $\mY$ to basis $\mZ$.
The logarithmic map can be replaced with the \textit{difference of projectors} $\mZ \mZ^T - \mY \mY^T$, which is a first-order approximation of $\log_{\mY}(\mZ) \mY^T + \mY \log_{\mY}(\mZ)^T$ -- a linear isometry of $\log_{\mY}(\mZ)$.\footnote{$||\log_{\mY}(\mZ) \mY^T + \mY \log_{\mY}(\mZ)|| = \sqrt{2}||\log_{\mY}(\mZ)||$ for any Schatten norm, e.g., Frobenius. Proof omitted for brevity.}
The difference of projectors can be favored over the logarithmic map because, like the chordal distance, only matrix multiplies are required instead of an SVD.



Our goal is to apply elementary manifold operations (and their approximations) to define and track motion on the Grassmannian. We draw analogies to tracking Newtonian motion in Euclidean space.

\section{Methods}
\vspace*{-5pt}

The typical optimization problem for subspace estimation or low-rank approximation is:
\begin{equation}\label{eq:lora}
\argmin_{\mY\in G(n,d)} \|\mX - \mY\mY^T\mX\|_F^2
=\argmin_{\mY\in G(n,d)} C(\mY,\mX)
\end{equation}
where $\mX\in\R^{n\times N}$ is a collection of $N$ data samples and $\mY\mY^T\mX$ is the projection of $\mX$ onto the span of $\mY$.
Solving the problem finds the $d$-dimensional subspace which best represents data from an $n$-dimensional ambient space.
Note that this is a non-convex optimization problem, but nonetheless a closed form minimizer is computed using the singular value decomposition (SVD) of the data truncated to the first $d$ left singular vectors.


Taking the SVD of a large dataset assumes the underlying subspace is constant --- an assumption which is violated in some applications such as array processing. We can instead consider a scenario where data arrives in batches $[\vx_{t,1}; \hdots; \vx_{t,B}]=\mX_t\in\R^{n\times B}$ for batch size $B$ with the goal of fitting one subspace $\mY_t$ to each batch. We will consider optimization problems of the form:
\begin{equation}\label{eq:reg_loss}
\argmin_{\{\mY_t\in G(n,d)\}}
\sum_t C(\mY_t, \mX_t)
+\lambda F(\{\mY_t\})
\hspace{5pt}
\end{equation}
The summation measures how well each subspace estimate $\mY_t$ captures the batch $\mX_t$ similarly to \eqref{eq:lora}. The regularization function $F$ (scaled by $\lambda$)
incentivizes the trajectory $\{\mY_t\}_{t=1}^T$ to follow a model for expected subspace motion, and the tunable hyperparameter $\lambda$ enables a tradeoff between matching either the data or the model. Finding a closed form minimizer to the above regularized least squares (RLS) problem is not possible in general, but if $F$ is differentiable, a gradient can be computed for each $\mY_t$ and used for gradient descent \cite{boumal_introduction_2023}.


\vspace*{-5pt}
\subsection{Dynamical Models on the Grassmannian}
\vspace*{-5pt}


The evolution of a subspace trajectory on the Grassmannian in discrete time can be represented by a dynamical model. Given some state $[\mY_t,~\hdots]$ which is the current subspace position plus any additional auxiliary variables, a dynamical model produces the one step evolution of this system to $[\mY_{t+1},~\hdots]$. We can construct simple dynamical models which trace out smooth trajectories on the manifiold.
The simplest possible dynamical model for subspace motion is the static model: $\mY_{t+1}=\mY_t$.
A regularizer for this model can be constructed as:
\begin{equation}\label{eq:reg_pos}
F_p(\{\mY_t\})=\sum_t d(\mY_t,\mY_{t+1})^2
\end{equation}
where $d$ is some measure of distance between subspaces, such as the geodesic or chordal distance.
Regularizing with the distance between adjacent subspaces punishes any deviation from the static model --- \eqref{eq:reg_pos} is minimized when the $\mY_t$ are all identical and the learned subspaces are static over the batches. Equations \eqref{eq:reg_loss} and \eqref{eq:reg_pos} together encourage learning subspaces which both fit the data and do not vary significantly across batches. We use the subscript $F_p$ to denote this as position regularization or position RLS, and we will use $F_{p,G}$ and $F_{p,C}$ when $d(\cdot,\cdot)$ is the geodesic or chordal metric, respectively.





One drawback of the position regularizer \eqref{eq:reg_pos} is that the set $\{\mY_t\}$ is encouraged to stay close together, so tracking performance may suffer if there is non-negligible subspace motion between batches.
We can instead consider a dynamical model which acts on both subspace position $\mY_t$ and tangential velocity $\mH_t$: 
\begin{equation}\label{eq:dynamical_vel}
\begin{bmatrix}
    \mY_{t+1} \\[5pt] \mH_{t+1}
\end{bmatrix}
=
\begin{bmatrix}
    \exp_{\mY_t}(\mH_t) \\[5pt] \Gamma_{\mY_t}^{\mH_t}(\mH_t)
\end{bmatrix}
\end{equation}
The $\mY_t$ and $\mH_t$ represent similar concepts to position and velocity in Euclidean space; this dynamical model is analogous to a particle moving in a straight line with constant velocity (as opposed to constant position with the static model). The exponential map ensures position iterates stay on the manifold, and the parallel transport operator $\Gamma$ propagates a velocity vector from the tangent space at $\mY_t$ to the tangent space at $\mY_{t+1}$ \cite{edelman_geometry_1998}.
We can define a regularizer with the model \eqref{eq:dynamical_vel} and using the geodesic metric as follows:
\begin{equation}\label{eq:reg_vel}
F_{v,G}(\{\mY_t,\mH_t\})=
\sum_t \|\mH_{t+1} - \Gamma_{\mY_t}^{\mH_t}\mH_{t}\|_F^2
\end{equation}
The regularizer is minimized when tangential velocity is constant, i.e., the path taken by the $\mY_t$'s is a constant-speed geodesic.
We note that the extension to an acceleration-based model could be constructed, but it would require second-order motion on the Grassmannian, which is more intensive to compute \cite{hong_parametric_2015}.
We can make substitutions to express $F_{v,G}$ purely in terms of $\{\mY_t\}$ (assuming small enough $\mH_t$):
\begin{equation}
\mH_{t+1}=\log_{\mY_{t+1}}(\mY_{t+2})
,\hspace{5pt}
\Gamma_{\mY_t}^{\mH_t}\mH_t=-\log_{\mY_{t+1}}(\mY_{t})
\end{equation}
To save computation with the logarithms and avoid SVDs, we can make a first order chordal approximation to the logarithms using the difference of projectors as discussed in \ref{sec:formulation}.
\[
\|\mH_{t+1} - \Gamma_{\mY_t}^{\mH_t}\mH_{t}\|_F
\]
\begin{equation}
\approx\frac{1}{\sqrt{2}}\|\mY_{t+2}\mY_{t+2}^T - 2\mY_{t+1}\mY_{t+1} + \mY_{t}\mY_{t}^T\|_F
\end{equation}
The chordal velocity regularizer $F_{v,C}$ is then:
\begin{equation}\label{eq:reg_vel_approx}
\small
F_{v,C}(\{\mY_t\})=
\sum_t \|\mY_{t+2}\mY_{t+2}^T - 2\mY_{t+1}\mY_{t+1} + \mY_{t}\mY_{t}^T\|^2_F
\end{equation}

\vspace*{-5pt}
\subsection{Grassmannian Gradients}
\vspace*{-5pt}

To solve RLS problems like \eqref{eq:reg_loss} with position or velocity regularization, the gradients of functions on the Grassmannian need to be computed. Reference \cite{edelman_geometry_1998} defines the gradient of a function in Euclidean space constrained to the Grassmannian:
\begin{equation}
\nabla_{\mY} F = \Pi(T_{\mY}G)F_{\mY} = (\mId - \mY\mY^T)F_{\mY}
\end{equation}
which is the Euclidean gradient of the function $F_{\mY}$ projected onto the tangent space at $\mY$.
We briefly state the gradients needed for performing gradient descent on the RLS problem. The gradient derivations are straightforward, but we omit them for brevity.
The gradient of the batch loss term in \eqref{eq:reg_loss} is:
\begin{equation}\label{eq:grad_batch}
\nabla_{\mY_t} C(\mY_t, \mX_t)
=-2(\mId - \mY_t\mY_t^T)\mX_t\mX_t^T\mY_t
\end{equation}
The gradient of the position regularizer \eqref{eq:reg_pos} with the geodesic metric is:
\[
\nabla_{\mY_t}F_{p,G}(\{\mY_t\})
=\nabla_{\mY_t}\sum_t \|\log_{\mY_t}(\mY_{t+1})\|_F^2
\]
\begin{equation}\label{eq:grad_pg}
=-2(\log_{\mY_t}(\mY_{t+1}) + \log_{\mY_t}(\mY_{t-1}))
\end{equation}
Similarly for \eqref{eq:reg_pos} with the chordal metric:
\[
\nabla_{\mY_t}F_{p,C}(\{\mY_t\})
=\nabla_{\mY_t}\sum_t \|\mY_{t+1}\mY_{t+1}^T-\mY_t\mY_t^T\|_F^2
\]
\begin{equation}\label{eq:grad_pc}
=-4(\mId-\mY_t\mY_t^T)(\mY_{t-1}\mY_{t-1}^T + \mY_{t+1}\mY_{t+1}^T)\mY_t
\end{equation}
The gradient for the geodesic velocity regularizer \eqref{eq:reg_vel} is more involved, so we leave it to future work. The gradient of the chordal velocity regularizer \eqref{eq:reg_vel_approx} is:
\[
\nabla_{\mY_t} F_{v,C}(\{\mY_t\}) =
-4(\mId - \mY_t\mY_t^T)(4\mY_{t+1}\mY_{t+1}^T
\]
\begin{equation}\label{eq:grad_va}
+ 4\mY_{t-1}\mY_{t-1}^T - \mY_{t+2}\mY_{t+2}^T - \mY_{t-2}\mY_{t-2}^T)\mY_{t}
\end{equation}
For position and velocity regularizers, the gradient for $\mY_t$ near the edges of the sums will be missing terms.
Given a set of data batches $\{\mX_t\}$ and initial subspace estimates $\{\mY_t\}$, we can use \eqref{eq:grad_batch} along with either \eqref{eq:grad_pg}, \eqref{eq:grad_pc}, or \eqref{eq:grad_va} to perform gradient descent on \eqref{eq:reg_loss} for different regularizers. We take gradient steps on all $\mY_t$ simultaneously with learning rate $\alpha$ and iteration $k$: $\mY_t^{(k+1)}=\exp_{\mY_t^{(k)}}(-\alpha\nabla_{\mY_t^{(k)}}(C+F))$. Gradient descent on manifolds is well-studied and discussed in \cite{boumal_introduction_2023}.
\vspace*{-5pt}
\section{Results}
\vspace*{-5pt}

\begin{figure}
\centering
\includegraphics[width=2.7in]{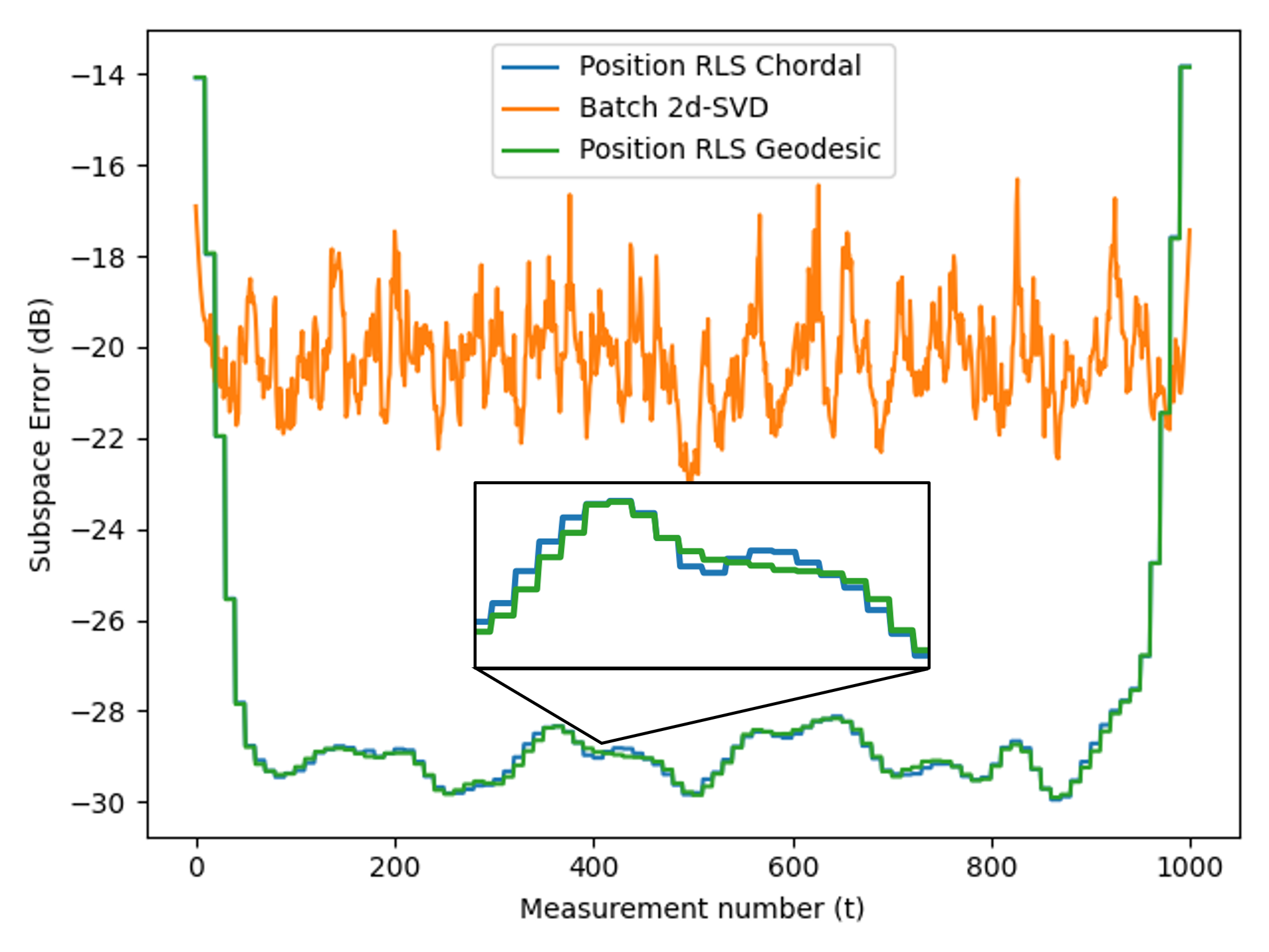}
\caption{A comparison between chordal and geodesic position RLS on the Grassmannian. The underlying subspace is drawn from a geodesic on the manifold with small tangential velocity. Subspace error is computed as the geodesic distance between the true and estimated subspaces. Both formulations achieve similar subspace tracking performance, but the geodesic incurs a nearly $5\times$ increase to runtime.}
\label{fig:rls_pos_performance}
\vspace*{-5pt}
\end{figure}

Using the developed RLS framework, we aim to accurately track changing subspaces from noisy measurements. We will briefly discuss computational aspects as well.
We will first compare the performance of geodesic and chordal versions of position RLS by using \eqref{eq:grad_batch} with \eqref{eq:grad_pg} or \eqref{eq:grad_pc} for gradient descent.
Underlying subspaces are generated with a random geodesic on the Grassmannian $G(64, 10)$: $\mY_t^{\star}=\exp_{\mY_0^{\star}}(t\mH_0^{\star})$ with $\|\mH_0^{\star}\|_F=10^{-2}$.
The data are generated as $\vx_t = \mY_t^{\star}\valpha_t + \veta_t$ for basis coefficients $\valpha_t\sim N(0,\mId)$, noise $\veta_t\sim N(0, \sigma^2\mId)$, and noise level $\sigma=10^{-2}$.
Subspace error is measured as the geodesic distance between learned $\mY_t$ and true $\mY_t^{\star}$.
The regularization coefficient is chosen to be $\lambda=1000$.
We use a batch size of $B=10$ and perform gradient descent with a constant learning rate of $\alpha=10^{-5}$ for 100 iterations. We take windowed $2d$-SVDs of the measurements for comparison (since $1d$-SVDs are noisier) and use them to initialize the RLS algorithms. The runtimes for gradient descent are 15.287 seconds for geodesic $F_{p,G}$ and 3.225 seconds for chordal $F_{p,C}$ -- a speedup of $4.74\times$. As discussed in \ref{sec:formulation}, the runtime reduction stems from the usage of SVDs in the computation of logarithmic maps for the geodesic loss \eqref{eq:grad_pg}, whereas only matrix multiplies are required for the chordal loss \eqref{eq:grad_pc}.
As shown in Fig. \ref{fig:rls_pos_performance}, the computational savings using the chordal metric do not incur a tracking penalty.
Subspace tracking performance at both edges of the batches suffers due to lack of data.

\begin{figure}
\centering
\includegraphics[width=2.7in]{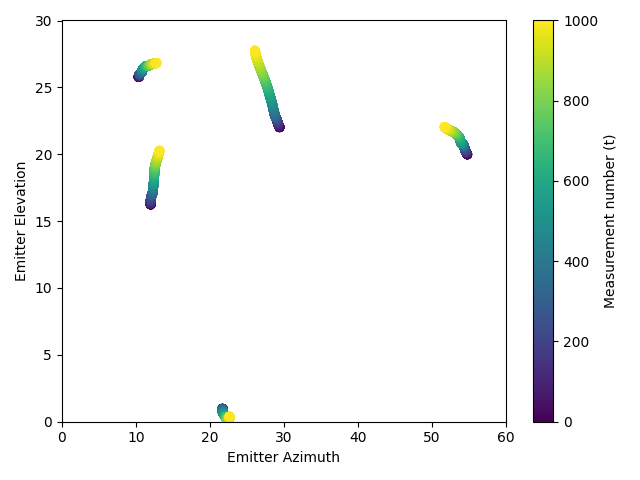}
\includegraphics[width=2.7in]{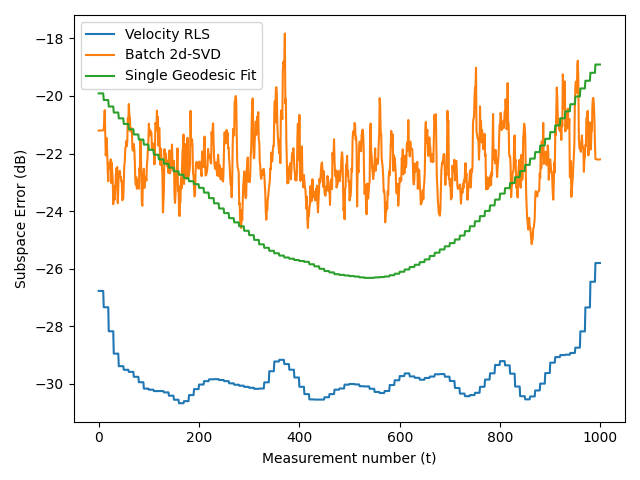}
\caption{Results for applying chordal velocity RLS to an antenna beamforming scenario. The scenario consists of random motion of 5 narrowband emitters in front of an $8\times8$ antenna array.
(a) The motion of narrowband emitters in azimuth and elevation over the course of the scenario.
(b) Velocity RLS tracking performance compared to batch $2d$-SVD, single geodesic fit \cite{blocker_dynamic_2023}, and noise level.
}
\label{fig:manifold_rls_results}
\vspace*{-5pt}
\end{figure}

We additionally applied chordal velocity RLS ($F_{v,C}$ from \eqref{eq:reg_vel_approx} as a regularizer in \eqref{eq:reg_loss}) to an antenna array processing scenario. The gradient descent hyperparameters are chosen identically as above. In the scenario, 5 narrowband emitters with random motion move in front of an $8\times8$, half-wavelength spaced antenna array as shown in Fig. \ref{fig:manifold_rls_results}a. The underlying subspace is given by:
\setlength{\abovecaptionskip}{-5pt}
\setlength{\belowcaptionskip}{-10pt}
\begin{equation}
\small
\mY^{\star}_t = \textrm{orth}([\Re{\mU^{\star}_t}; \Im{\mU^{\star}_t}])
\end{equation}
\begin{equation}
\mU^\star_t=
\small
\begin{bmatrix}
    \exp(j\vr_1\cdot\vk_1(t)) & \hdots & \exp(j\vr_1\cdot\vk_5(t)) \\
    \vdots & \ddots & \vdots \\
    \exp(j\vr_{64}\cdot\vk_1(t)) & \hdots & \exp(j\vr_{64}\cdot\vk_5(t)) \\
\end{bmatrix}
\end{equation}
Where $\vr_i$ are the receiver positions and $\vk_j$ are the wavevectors of the incident signals. We split the subspaces into real and imaginary parts to get a subspace rank of $d=10$.
Subspace tracking results for the array processing scenario are shown in Fig. \ref{fig:manifold_rls_results}b.
In addition to the $2d$-SVD, we also compare against fitting a single geodesic as is done in \cite{blocker_dynamic_2023}. As shown, the subspace dynamics are only partially captured by a single geodesic. However, velocity RLS is able to encode the disparate motion of narrowband emitters onto a smooth trajectory on the Grassmannian and achieve consistent subspace tracking performance. 

\vspace*{-9pt}
\section{Conclusion}
\vspace*{-9pt}
In this work, we have developed a regularized least squares (RLS) framework for subspace tracking on the Grassmannian. Dynamical models are constructed which trace smooth subspace trajectories on the manifold.
We apply chordal computations in place of geodesic to reduce runtime without deteriorating subspace tracking performance.
The velocity RLS algorithm is shown to capture complicated signal dynamics in a narrowband beamforming scenario.
\vfill

\pagebreak
\bibliographystyle{IEEEtran}
\bibliography{references}

\end{document}